\documentclass[twocolumn]{jpsj2} 
\title{Quantum Numbers for Excitations of Bose-Einstein Condensates\\in 1D Optical Lattices}

\author{Tomoya \textsc{Isoshima}$^{1,2}$ and Martti M. \textsc{Salomaa}$^{1,3}$}

\inst{$^{1}$Materials Physics Laboratory,
Helsinki University of Technology,\\
P.O.Box 2200 (Technical Physics), FIN-02015 HUT, Finland \\
$^{2}$Institute of Physics, Academia Sinica, Nankang, Taipei 11529, Taiwan\\
$^{3}$Department of Physics, Kinki University, Higashi-Osaka 577-8502}

\abst{
The excitation spectrum and the band structure
of a Bose-Einstein condensate in a periodic potential are investigated.
Analyses within full 3D systems,
finite 1D systems, and ideal periodic 1D systems are compared.
We find two branches of excitations in the spectra of the finite 1D model.
The band structures
for the first and (part of) the second band are
compared between a finite 1D and the fully periodic 1D systems,
utilizing a new definition of a effective wavenumber and a phase-slip number.
The upper and lower edges of the first gap coincide well between the two cases.
The remaining difference is explained by the existence of the two branches
due to the finite-size effect.
}

\kword{optical lattice, Bose-Einstein condensate, wavenumber, excitation spectra}

\begin{document}
\maketitle

\section{Introduction}

\begin{figure}[hbt]
\begin{center}
\includegraphics[width=7.5cm,clip]{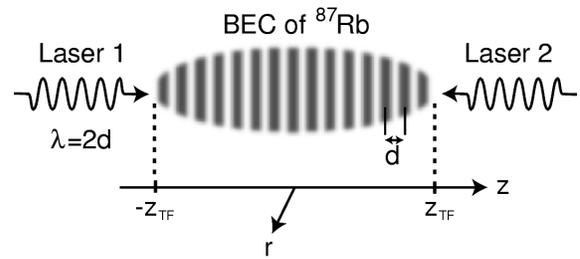}
\end{center}
\caption{
BEC of Rb atoms is confined along the $z$ and $r$ directions
by harmonic traps with frequencies $\nu_z$ and $\nu_r$ respectively.
The condensate consists of $5\times 10^4$ atoms of $^{87}$Rb.
Counterpropagating laser beams of wavelength $\lambda$ form a potential
of period $d=\lambda / 2$. 
}
\label{fig:schematic}
\end{figure}

Bose-Einstein condensate (BEC) of the atomic gases trapped in the periodic potential of 
an optical lattice are currently intensively
investigated.~\cite{pitaevskii-stringari,scott,plata}
These systems have been researched also
as the candidates of quantum computing.~\cite{zengbing,katharina}
The condensate is usually confined with a potential generated by a magnetic field,
and also modulated by the periodic potential formed by the counterpropagating laser beams;
for a schematic illustration of the experimental arrangement, see Fig.~\ref{fig:schematic}. 

The system may be treated as periodic~\cite{pitaevskii-stringari,sorensen}
ignoring the confinement.
Within this periodic framework,
the wavenumber of the excitations becomes periodic and the whole spectrum
splits into band structures which are similar to those within solid-state physics.
Conversely, in the limit of a weak periodic potential,
the condensate reduces into a harmonically trapped, cigar-shaped condensate.
The classifications of the collective excitations are 
closely related to the symmetry of the confining trap.
Angular momenta and other indices in spherically symmetric systems
have been used.~\cite{pethick-smith}
Within cigar-shaped systems,
the wavenumber along the $z$-axis~\cite{mizushima} is also
a useful quantum number to characterize an excitation.

In this paper, we calculate the full excitation spectra for Bose-Einstein condensates
of atomic gases trapped in a periodic optical potential and
a harmonic confinement potential. 
We aim to classify and understand the excitation spectra properly,
using the similarity with the purely periodic systems and the cigar-shaped systems.

\section{Condensate in 1D Optical Lattices}

We study the excitation spectra of Bose-Einstein condensates (BEC)
confined in a composite potential
\begin{eqnarray}
    V(r,z) &=&
         \frac{m}{2} (2\pi\nu_r)^2 r^2
       + V_z(z)
       + V_\mathrm{opt}(z),
\label{eq:v}
\\
    V_z(z) &=& \frac{m}{2} (2\pi\nu_z)^2 z^2, 
\label{eq:vz}
\\
    V_\mathrm{opt}(z) &=& s E_\mathrm{R} \sin^2 \left( \pi \frac{z}{d} \right)
\label{eq:vopt}
\end{eqnarray}
that consists of an optical lattice in the principal direction of consideration 
and harmonic confinement potentials along the lateral dimensions.
Here 
$s$ is an intensity parameter of the periodic optical potential,
$E_\mathrm{R} = \frac{\hbar^2}{2m}\left(\frac{\pi}{d}\right)^2$ is a recoil energy,
$m$ is the mass of a $^{87}$Rb atom, $\nu_r$ and $\nu_z$ are the frequencies of the
harmonic potentials, and $d$ is the period of the potential.
We employ $d=0.395 \, \mu\mathrm{m}$ which is one-half of
the typical~\cite{scott}
wavelength $\lambda$ for the laser beams, $\nu_r = 100 \, \mathrm{s}^{-1}$
and $\nu_z = 30 \, \mathrm{s}^{-1}$ which are also within the typical ranges 
in experiments.
A condensate with 3D geometry, 1D geometry and
a periodic 1D situation are simulated
and compared
in order to determine the effects of the confinement potentials.

\begin{figure}[hbt]
\begin{center}
\includegraphics[width=7.5cm,clip]{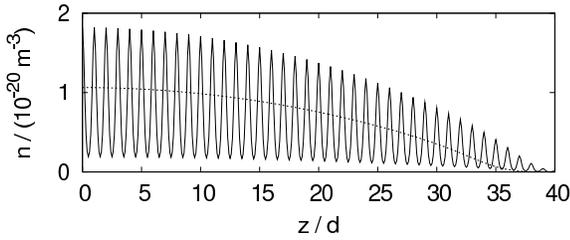}
\end{center}
\caption{Solid curve: the condensate density $|\phi(r,z)|^2$ at
 $r=0$ with the lattice intensity $s=5$.
The range $z \ge 0$ is plotted.
The dotted line is the density profile in the absence of the optical lattice ($s=0$).}
\label{fig:contour}
\end{figure}

The BEC in an optical lattice with the lower value of 
the intensity $s$
is well described by the Gross-Pitaevskii equation~\cite{pethick-smith}
\begin{equation}
    \left( - C \nabla^2 + V(r,z) + g|\phi(r,z)|^2 \right)\phi(r,z) = \mu\phi(r,z)
\label{eq:gp}
\end{equation}
where $\phi(r,z)$ is the wavefunction of the condensate,
$\mu$ is the chemical potential,
$g = \frac{4\pi\hbar^2a}{m}$ is an interaction parameter
and $C = \hbar^2/(2m)$ is constant.
We treat a condensate with $5\times 10^4$ atoms of $^{87}$Rb.
The scattering length $a = 5.4 \, \mathrm{nm}$ 
is employed for the $^{87}$Rb atoms.
The period of the optical lattice is $d=0.395 \, \mu\mathrm{m}$ and
one lattice site is occupied by at most 1300 atoms.
The condensate occupies about 80 sites.
Figure \ref{fig:contour} shows the density $|\phi|^2$ of a typical condensate along the $z$-axis.

The condensate supports excitation spectra within the Bogoliubov framework.
The Bogoliubov equations~\cite{pethick-smith}
%
%
\begin{eqnarray}
    \left(- C\nabla^2 + V(r,z) + 2g|\phi|^2 - \mu\right)u_q
        - g\phi^2v_q \!\!&=&\!\! \varepsilon_q u_q,
\nonumber \\
\label{eq:bdg1}
\\
    \left(- C \nabla^2 + V(r,z) + 2g|\phi|^2 - \mu\right)v_q
        - g\phi^{\ast 2}u_q \!\!&=&\!\! -\varepsilon_q v_q,
\nonumber \\
\label{eq:bdg2}
\end{eqnarray}
%
%
yield the excitation energies $\varepsilon_q$
and the corresponding wavefunctions, $u_q$ and $v_q$.
The condensate and the excitations may have finite angular momenta
which are essential to treat the systems having vortices~\cite{martikainen}.
We only consider excitations having zero angular momentum and concentrate on
the wavenumbers along the principal $z$-axis in this article.

\begin{figure}[tb]
\begin{center}
\includegraphics[width=7.5cm,clip]{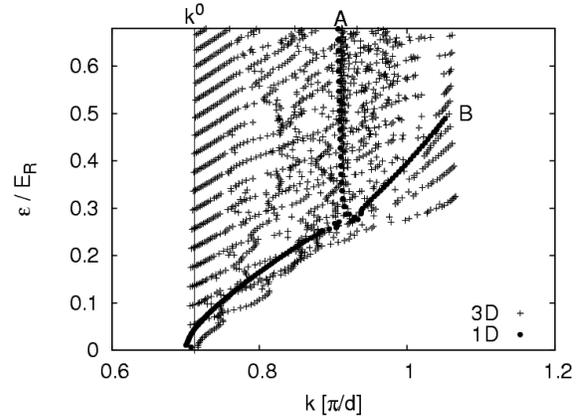}\\
(a)\\

\vspace{1\baselineskip}
\includegraphics[width=7.5cm,clip]{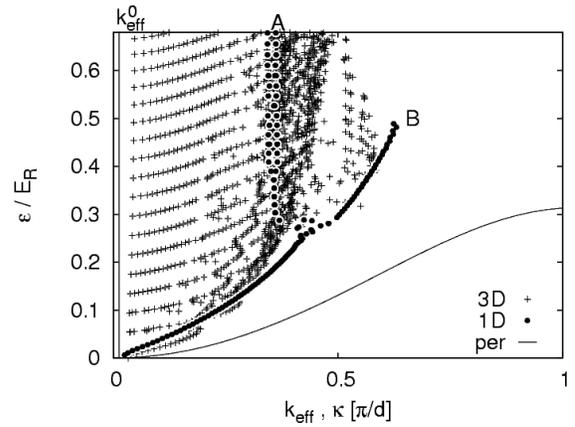}\\
(b)\\
\end{center}
\caption{Excitation spectra within 3D analysis (crosses), 
finite-size 1D model (bullets) and periodic 1D analysis (solid line).
(a) The wavenumber $k$ defined by Eq.~(\ref{eq:k}); and
(b) the effective wavenumber $k_\mathrm{eff}$ as given in Eq.~(\ref{eq:keff}). 
Both 1D spectra (bullets) feature two distinct branches,
that are named branch A and B.
The patterns of the 3D spectra significantly differ
between the right- and left-hand sides for branch A.
The range of $k_\mathrm{eff}$ does not depend on $s$.
The vertical lines in (a) and (b) indicate $k^0$ and $k^0_\mathrm{eff}$ 
respectively.
}
\label{fig:spectra}
\end{figure}

\section{Wavenumbers}

In the absence of an optical lattice potential, 
the condensate has a cigar-like shape and
the wavenumber of an excitation may be defined through
differentiating the wavefunction.
In contrast, 
we may consider an ideal periodic 1D system to
treat only the effects of the 1D optical lattice potential and
ignore the confinement potential. 
Within the periodic 1D system~\cite{pitaevskii-stringari},
the wavefunction may be expressed in the form 
\begin{equation}
    \phi_{\kappa}(z) = \exp( \mathrm{i} \kappa z ) u_{\kappa}(z),
\label{eq:quasi_k}
\end{equation}
where $\kappa$ is a quasi-wavenumber.

In order to consider these two aspects of this system,
we introduce the two definitions for wavenumbers
\begin{equation}
  k \equiv
        \sqrt{\left|\int u^\ast \frac{\partial^2}{\partial z^2} u \,\mathrm{d}\mathbf{r}\right|
       {\LARGE /} \int|u|^2  \,\mathrm{d}\mathbf{r}},
\label{eq:k}
\end{equation}
and
\begin{equation}
  k_{\mathrm{eff}}(d) \equiv
    \sqrt{
    \frac{-\sum_j u^{\ast}_j \left(u_{j-1} + u_{j+1} -2 u_{j} \right)}{d^2 \sum_j |u_j|^2}},
\label{eq:keff}
\end{equation}
where $u_j$ is the value of the wavefunction $u(r,z)$
at the center of the $j$-th site.
The definition of $k_{\mathrm{eff}}$ arises from
the discretization of the second derivative in eq.~(\ref{eq:k}).
Hence the two wavenumbers $k$ and $k_{\mathrm{eff}}$ coincide in the limit as $d \to 0$.
The $k_{\mathrm{eff}}$ is compatible with the quasi-wavenumber $\kappa$
within a periodic system for $|\kappa| \ll 1/d$.
The corresponding wavenumbers of the condensate are
\begin{eqnarray}
    k^{0} &\equiv&
        \sqrt{\left|\int \phi^\ast \frac{\partial^2}{\partial z^2} \phi \,\mathrm{d}\mathbf{r}\right|
       {\LARGE /} \int|\phi|^2  \,\mathrm{d}\mathbf{r}},
\label{eq:k0}
\\
  k^{0}_{\mathrm{eff}} &\equiv&
    \sqrt{
    \frac{-\sum_j \phi^{\ast}_j \left(\phi_{j-1} + \phi_{j+1} -2 \phi_{j} \right)}{d^2 \sum_j |\phi_j|^2}}.
\label{eq:k0eff}
\end{eqnarray}

Figures \ref{fig:spectra}(a) and \ref{fig:spectra}(b) represent the excitation
spectra using $k$ and $k_{\mathrm{eff}}$, respectively.
Note, in particular, that the wavenumbers $k$ are strongly shifted
and modified with respect to
those ($|\kappa| \le \pi/d$) 
of an infinite periodic 1D system.~\cite{pitaevskii-stringari}
The wavenumbers $k$ are appreciably high compared to $\pi/d$.
This is because the intensity of the wavefunction $u$ oscillates with the period $d$
along the length $z$.
The effective wavenumber $k_\mathrm{eff}$ defined in eq.~(\ref{eq:keff}) 
is not explicitly modified by this oscillation.
Its value does not exceed $\pi/d$.
The lowest $k_\mathrm{eff}$ value is close to 0, which is the minimum $|\kappa|$
in the purely periodic system.

The alignments of the excitations in the plots become disordered [Fig.~\ref{fig:spectra}(b)]
and dense [Fig.~\ref{fig:spectra}(b)] above certain threshold values
of $k$ and $k_\mathrm{eff}$.
The spectra (the bullets in Figs.~\ref{fig:spectra}(a) and \ref{fig:spectra}(b)) of
the corresponding 1D model systems make the change in the alignments plausible.
The 1D model system is calculated by treating only the $r=0$ case
in eqs.~(\ref{eq:v} -- \ref{eq:bdg2}).
The chemical potential $\mu$ is the same as that of a corresponding 3D system,
such that the particle densities are close to those of a 3D system at $r=0$.

The 1D system features two branches in the spectra, which are named A and B.
The spectra in the 3D analysis differ considerably
between the left- and the right-hand sides of the branch A in Fig.~\ref{fig:spectra}(b).
The branch A explains the dense population of the excitations
around 
$k = 0.9 \pi/d$ (hidden behind the bullets in Fig.~\ref{fig:spectra}(a))
and
$k_\mathrm{eff} = 0.5 \pi/d$
in the spectra of the 3D system.
The 3D systems have modes corresponding to the branch A with
various radial wavenumbers in the narrow range of the wavenumber $k_\mathrm{eff}$
of principal direction.

The vertical (\textit{i.e.}, having definite $k$ and $k_\mathrm{eff}$ values)
branch A does not exist within the purely periodic model.  
For comparison,
we consider spatially periodic systems~\cite{pitaevskii-stringari}.
The condensate wavefunction has the same form as in eq.~(\ref{eq:quasi_k})
and $u_{\kappa}(z)$ obeys the wave equation
\begin{eqnarray}
\lefteqn{
    -C \left(\frac{\mathrm{d}}{\mathrm{d}z} - \mathrm{i}\kappa\right)^2 u_\kappa(z)
}\nonumber \\ &&
    + \left(g|u_\kappa(z)|^2 + V_\mathrm{opt}\right)u_\kappa(z) = \mu(\kappa) u_\kappa(z)
\label{eq:3dpot}
\end{eqnarray}
where $-\frac{\pi}{d} \le \kappa \le \frac{\pi}{d}$.
The potential energy $V_\mathrm{opt}$ is the last term in eq.~(\ref{eq:v}).
The period of the system is $d$ and
the average density of atoms is $10^{20} \, \mathrm{m}^{-3}$,
which is comparable to the densities within the previous 3D and 1D analyses.

The excitation energies are written using the energy of a condensate wavefunction 
\begin{eqnarray}
\lefteqn{
E_\kappa = \int
        -C u_\kappa^{\ast}\left(\frac{\mathrm{d}}{\mathrm{d}z}
        - \mathrm{i} \kappa \right)^2 u_\kappa(z)
}\nonumber \\ &&
        + \frac{g}{2}|u_\kappa(z)|^4 
        + V_\mathrm{opt}|u_\kappa(z)|^2 
    \mathrm{d}z. 
\end{eqnarray}
The solid line in Fig.~\ref{fig:spectra}(b) represents the spectra
$\varepsilon_\kappa = (E_\kappa - E_0)/N$,
which are comparable to the previous spectra that utilize $k_\mathrm{eff}$.

\begin{figure}
\begin{center}
\includegraphics[width=7.5cm,clip]{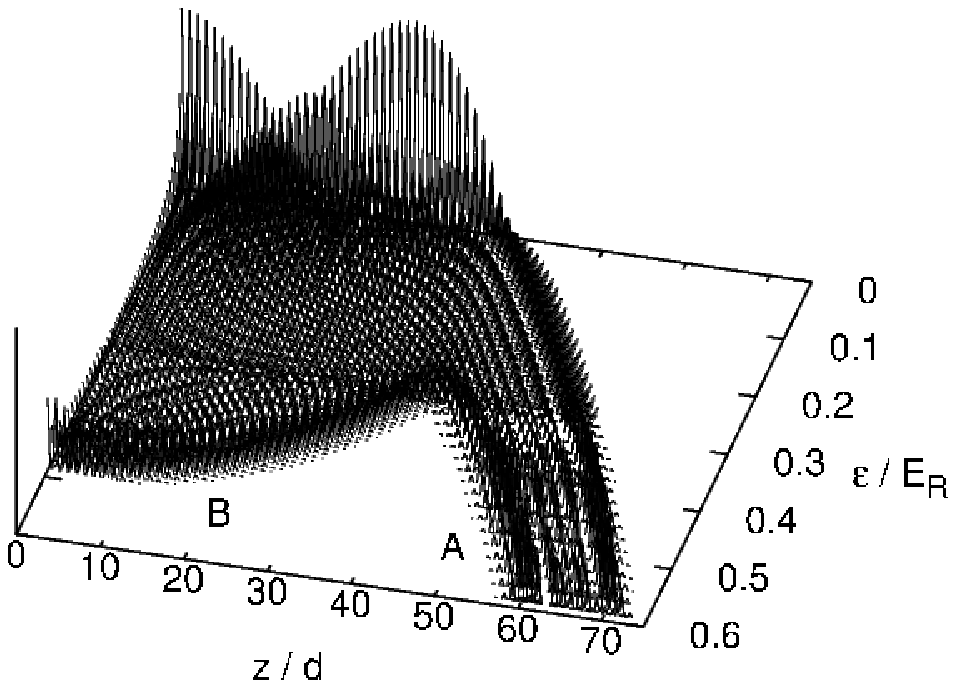}\\
(a)\\

\vspace{1\baselineskip}
\includegraphics[width=7.5cm,clip]{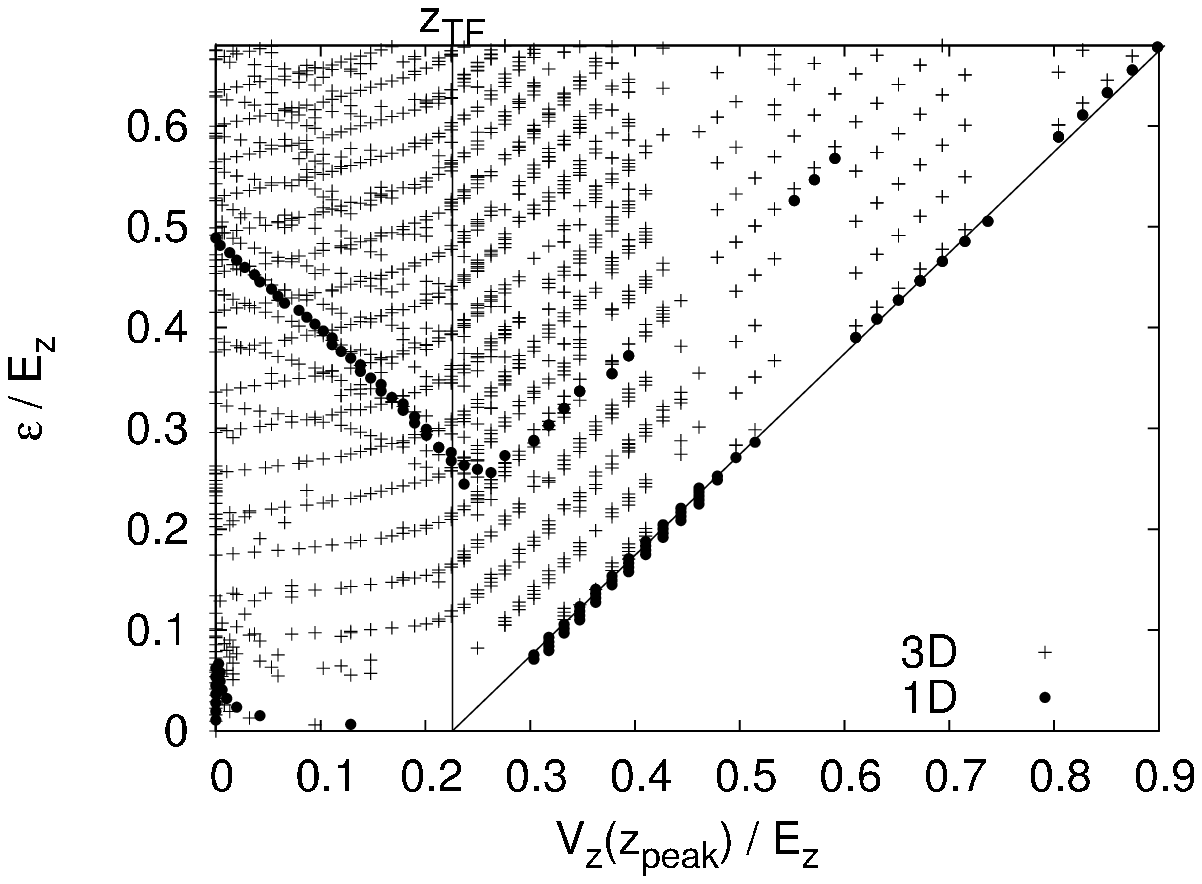}\\
(b)\\

\vspace{1\baselineskip}
\includegraphics[width=7.5cm,clip]{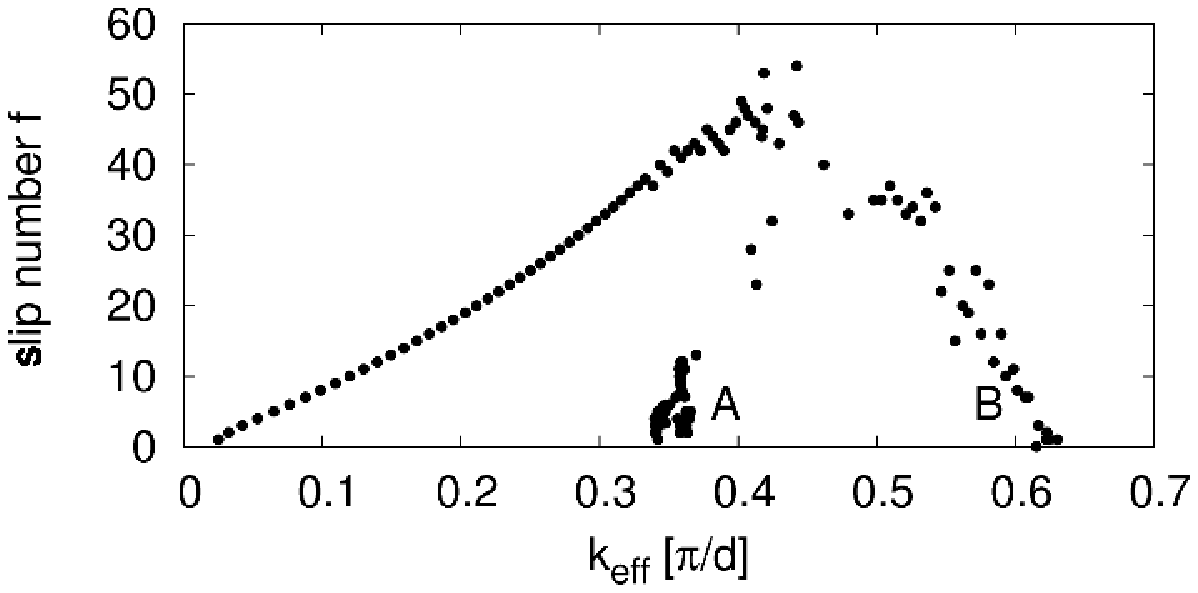}\\
(c)
\end{center}
\caption{
(a) Intensities of the excitations $|u(z)|^2 + |v(z)|^2$ within the 1D framework.
There is a spatial separation between the excitations of the two branches A and B
in Figs.~\ref{fig:spectra}(a) and \ref{fig:spectra}(b).
(b) Excitation energy vs potential energy
$V_z(z) = \frac{m}{2} (2\pi\nu_z)^2 z^2$
at the peak values of the amplitude of wavefunctions $|u|^2$.
The relation is linear beyond ($z_{\mathrm{peak}} > z_\mathrm{TF}$) the condensate,
both in the 3D and within the 1D framework.
Here $z_\mathrm{TF} = 14.65 \,\mu \mathrm{m}$ is a half the Thomas-Fermi length of the
condensate for $s=0$.
The vertical line indicates $V_z(z_\mathrm{TF})$.
The slope of the solid oblique line is 1.
(c)
The number of the phase slips $f$.
Labels A and B indicate the two branches.
The smooth increase in $f$ breaks down at the branch separation.
The intensity $s=5$ was used in these computations.
}
\label{fig:branches}
\end{figure}

\section{Origin of the Branches}

Figure \ref{fig:spectra}(b) shows that 
branch A does not exist in the spectra of the purely periodic system.
In order to explain the origin of branch A, we investigate 
the profiles of the wavefunctions within the 1D system.
Figure \ref{fig:branches}(a) shows the spectral evolution within the 1D analyses.
Excitations in the branch A are localized
outside of the condensate $|z| >  z_\mathrm{TF}$.
Here $z_\mathrm{TF} = 14.6 \, \mu\mathrm{m}$ is one-half the condensate length within the
Thomas-Fermi approximation in the absence of the periodic potential. 
The Bogoliubov eqs.~(\ref{eq:bdg1}) and (\ref{eq:bdg2}) reduce to a Schr\"odinger
equation outside the condensate ($|\phi| \to 0$).  
The graph of $V_z( z_\mathrm{peak} )$ in Fig.~\ref{fig:branches}(b) shows 
that the increase of the excitation energies
in branch A simply depends on the $z$-confinement potential $V_z( z_{\mathrm{peak}} )$,
where $z_{\mathrm{peak}}$ indicates the peak intensity of an excitation wavefunction $u_q(z)$.
The regular spacing of the excitation energies $\varepsilon$ may be
understood in terms of the $z$-confinement trap and
the period $d$ of the optical lattice.
The energy spacing $\Delta\varepsilon$ reflects the $z$-confinement potential since
\begin{equation}
    \Delta\varepsilon \simeq V(z+d) - V(d).
\label{separations}
\end{equation}
Excitations within the 3D framework in Fig.~\ref{fig:branches}(b) have the same features.
Branch A in the 1D systems corresponds to excitations localized
outside the condensate in 3D systems.

The origin of the branch B is clearly seen through defining a number $f$ of phase slips.
Unlike for the purely periodic system, the differences in
the complex phases of a wavefunction $u_q(z)$ 
between the neighboring sites are always either 0 or $\pi$.
Therefore, we may define a phase-slip number
\begin{equation}
f \equiv \sum_j \frac{1}{2} \left| \frac{u_{j}}{|u_j|} - \frac{u_{j-1}}{|u_{j-1}|}\right|
\end{equation}
for each of the excitations.
Figure \ref{fig:branches}(c) shows this number $f$ as a function of the 
effective wavenumber, $k_\mathrm{eff}$.
The phase-slip number $f$ increases as $k_{\mathrm{eff}}$ increases
from zero towards the branch separation.
In other words, the peak of the phase-slip number separates the branch B
from the rest of the excitations.

\section{Band Gap}

\begin{figure}
\begin{center}
\includegraphics[width=7.5cm,clip]{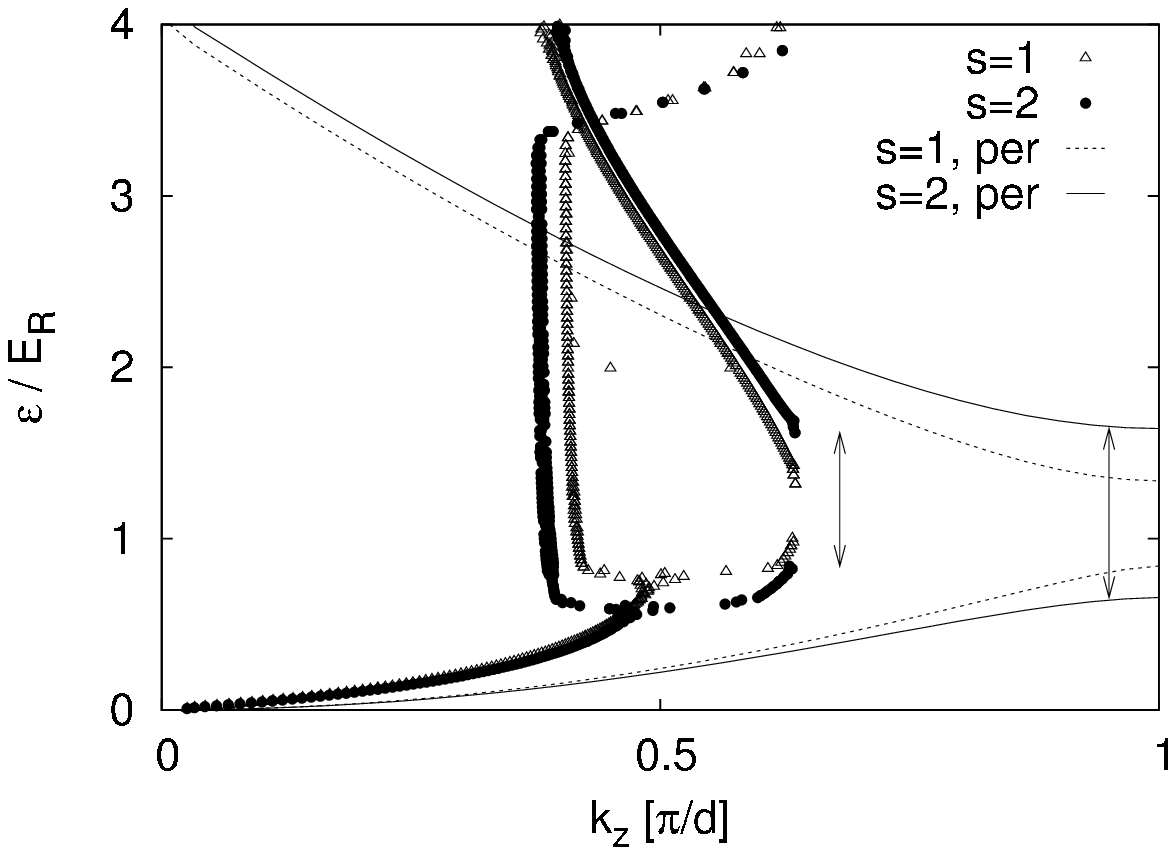}\\
(a)\\

\vspace{1\baselineskip}
\includegraphics[width=7.5cm,clip]{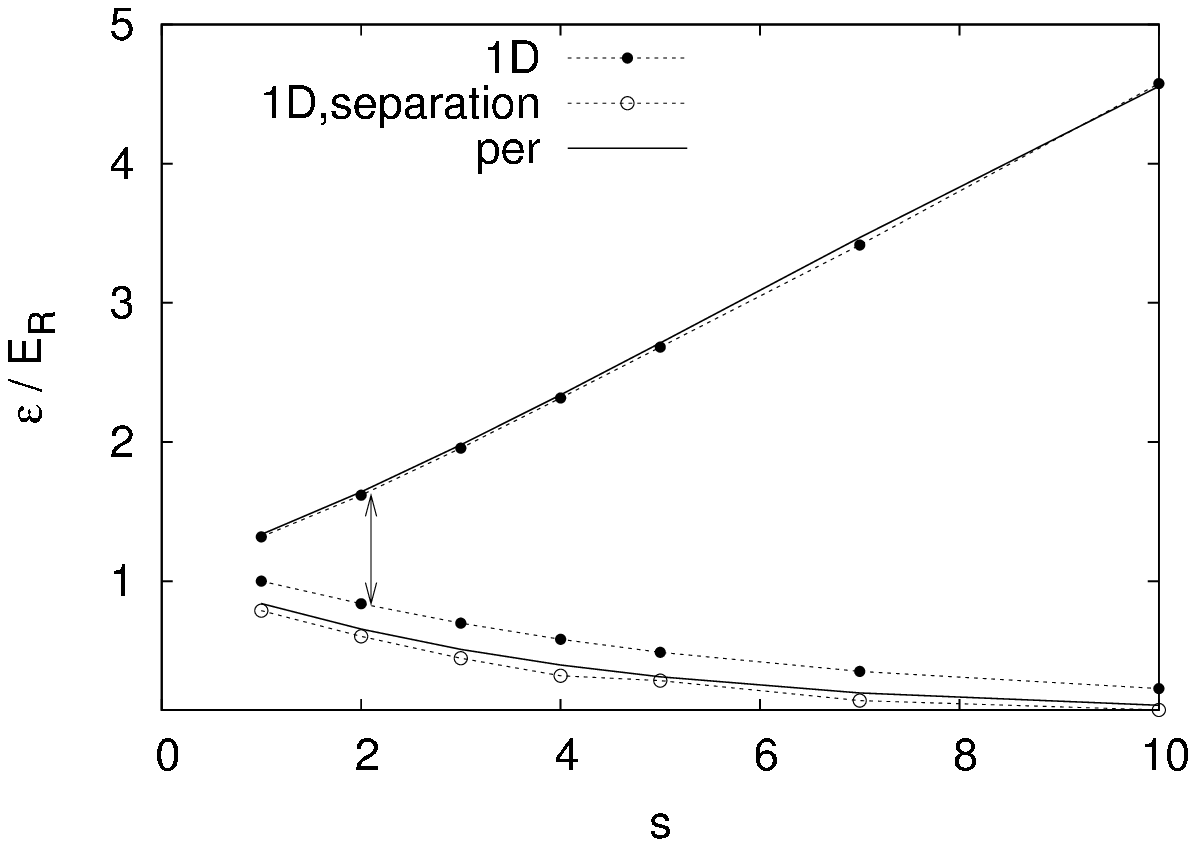}\\
(b)
\end{center}
\caption{
(a) Excitation spectra of the 1D system and the periodic system
for the optical lattice intensities $s=1$ and $2$.
Both spectra display a band gap indicated by the arrows.
(b) The gap between the first and second bands is plotted as a function of $s$.
The solid lines are the upper and lower edges of the gap within the periodic system.
The bullets are those for 1D systems.
The open circles show the excitation energies of wavefunctions
which have the maximum number $f$ of phase-slips for each $s$.
}
\label{fig:gap}
\end{figure}

The 1D systems feature gaps in the spectra if we ignore the branch A
as shown in Figure \ref{fig:gap}(a).
The range of the gap is close to the first gap in the corresponding periodic system.
The gaps between the first and the second bands are plotted
in Fig.~\ref{fig:gap}(b) as functions of $s$
for the range $1 \le s \le 10$ of intensities for the optical lattice.
The upper limit of the gap,
i.e., the upper solid line and the upper series of bullets in Fig.~\ref{fig:gap}(b)
coincides between the two systems.
There is difference of about $0.17 E_\mathrm{R}$ between the two lower limits,
which are the lower solid line and
the lower series of bullets in Fig.~\ref{fig:gap}(b).

The excitation energy of a wavefunction
which has the maximum number $f$ of phase slips 
indicates the separation of the branches.
The open circles in Fig.~\ref{fig:gap}(b) show the excitation energies at the separation.
These energies coincide better with the lower limit of the band in the periodic system.
The difference between the two lower limits emerges from the existence of the branch B.
Therefore, the two branches make a significant difference in the spectra 
between a finite system (1D) and an ideal periodic system.
Otherwise, the two spectra feature quite similar first bands and first band gaps.

\section{Summary}

We have investigated the excitation spectra and the band structures
of a Bose-Einstein condensate in periodic potentials.
Comparisons between calculations performed within 3D systems,
finite 1D systems, and ideal periodic 1D systems have been made.

Two branches are found in the spectra of the 1D model system.
The branch A, which makes a major difference between
the confined 1D system and a periodic 1D system, 
arises from the excitations that prevail outside the condensate.
Excitations in the other branch B
possess lower phase-slip numbers than the excitations
at the branch separation, despite the higher wavenumbers $k$ and $k_\mathrm{eff}$.
The structures corresponding to the branches are found 
in the spectra of the 3D system in Fig.~\ref{fig:branches}(b).

The band structure up to the beginning and along the 2nd band are
compared between the finite 1D system and the ideal periodic 1D system,
utilizing new definitions of a effective wavenumber and a phase-slip number.
The upper and lower edges of the first gap coincide well between the two systems.
The remaining differences are explained by the existence of branch B
in the excitation spectra of the 1D system.

The authors are grateful to
J.-P.~Martikainen and S. M. M. Virtanen for stimulating discussions
and to CSC-Scientific Computing Ltd (Espoo, Finland) for resources.
One of the authors (T.I.) is supported by the bilateral exchange program between the
Academy of Finland and JSPS, the Japan Society for the Promotion of Science; 
the other (MMS) is grateful to the JSPS for the award of a Visiting Fellowship in Japan.


\begin{thebibliography}{99}

\bibitem{pitaevskii-stringari}
L.~P.~Pitaevskii and S.~Stringari: Bose Einstein Condensation
(Oxford, Clarendon Press, 2003), Section 16.

\bibitem{scott}
R.~G.~Scott, A.~M.~Martin, S.~Bujkiewicz, T.~M.~Fromhold, N.~Malossi,
O.~Morsch, M.~Cristiani and E.~Arimondo: Phys. Rev. A \textbf{69} (2004) 033605.

\bibitem{plata}
J. Plata: Phys. Rev. A \textbf{69} (2004) 033604.

\bibitem{zengbing}
Z.-B. Chen and Y.-D. Zhang: Phys. Rev. A \textbf{65} (2002) 022318.

\bibitem{katharina}
K. Christandl and G. P. Lafyatis: e-print physics/0401041.

\bibitem{sorensen}
K.~B.-S{\o}rensen and K.~M{\o}lmer: Phys. Rev. A \textbf{58} (1998) 1480.

\bibitem{pethick-smith}
C.~J.~Pethick and H.~Smith: Bose-Einstein Condensation in Dilute Gases
(Cambridge University Press, 2002), Section 7.

\bibitem{mizushima}
T. Mizushima, M. Ichioka, K. Machida, T. Isoshima and M. M. Salomaa:
Laser Physics Vol. 14, No. 2 (2004) 295.

\bibitem{martikainen}
J.-P.~Martikainen and H.~T.~C.~Stoof: Phys. Rev. Lett. \textbf{91} (2003) 240403.

\end{thebibliography}
\end{document}